\documentclass[twocolumn,superscriptaddress,showpacs,preprintnumbers,amsmath,amssymb,a4paper]{revtex4}

\usepackage{amsmath,amssymb,bbm,epsfig}

\newcommand{\be}{\begin{equation}}
\newcommand{\ee}{\end{equation}}
\newcommand{\ba}{\begin{array}}
\newcommand{\ea}{\end{array}}
\newcommand{\bqa}{\begin{eqnarray}}
\newcommand{\eqa}{\end{eqnarray}}

\newcommand{\tr}{\mbox{Tr}}
\newcommand{\bra}[1]{\ensuremath{\langle #1 |}}
\newcommand{\ket}[1]{\ensuremath{| #1 \rangle}}

\usepackage{graphicx,color}
\definecolor{blue}{rgb}{0,0,0.8} 

\begin{document}

\title{Simulation of open quantum systems}

\author{Florian Mintert }
\affiliation{
Department of Physics, Harvard University,
17 Oxford Street, Cambridge Massachusetts, USA}
\affiliation{
Physikalisches Institut, Albert-Ludwigs Universit\"at Freiburg,
Hermann-Herder-Str. 3, Freiburg, Germany}
\author{Eric J. Heller}
\affiliation{
Department of Physics, Harvard University,
17 Oxford Street, Cambridge Massachusetts, USA}

\date{\today}

\begin{abstract}
We present an approach for the semiclassical treatment of open quantum systems.
An expansion into localized states allows restriction of a simulation to a fraction of the environment
that is located within a predefined vicinity of the system.
Adding and dropping environmental particles during the simulation
yields an effective reduction of the size of the system that is being treated.
\end{abstract}

\pacs{
02.60.Cb, 
02.70.Ss 
03.65.Sq 
03.65.Yz 
}

\maketitle

Any rigorous treatment of open system dynamics is impeded by the {\it de facto} impossibility
of simulating the dynamics of a system comprised of the system of interest and the typically macro-, or at least
mesoscopic environment.
Numerous prior approaches end up with {\em effective} equations of motion for the system itself,
where the dynamics of the environment is eliminated, such as a Lindblad equation
\cite{Lindblad:1976fj,hornberger:060601}
that describes the influence of an environment onto a system.
Such an effective theory, however, is always subject to approximations,
such as a Markov approximation, and/or taking the system-bath interaction, and/or the environment itself to be harmonic.

In this paper we describe an approach to open system dynamics based on the propagation of Gaussian states.
That is, an initial state is expanded in terms of Gaussian states,
and each of these components is then propagated individually.
Under the evolution of  a general Hamiltonian an initial Gaussian state will not remain Gaussian.
However, for well localized wave packets the Hamiltonian can be approximated to be locally harmonic,
so that the Gaussian character is preserved \cite{rick75,rick76,rick77}.
Such techniques have proven to accurately describe dynamical properties of various closed systems,
thermodynamic properties \cite{mandelshtam03,mandelshtam04},
and also open quantum systems \cite{fiete:022112,brodier:016204}.

\begin{figure}
\includegraphics[width=0.35\textwidth]{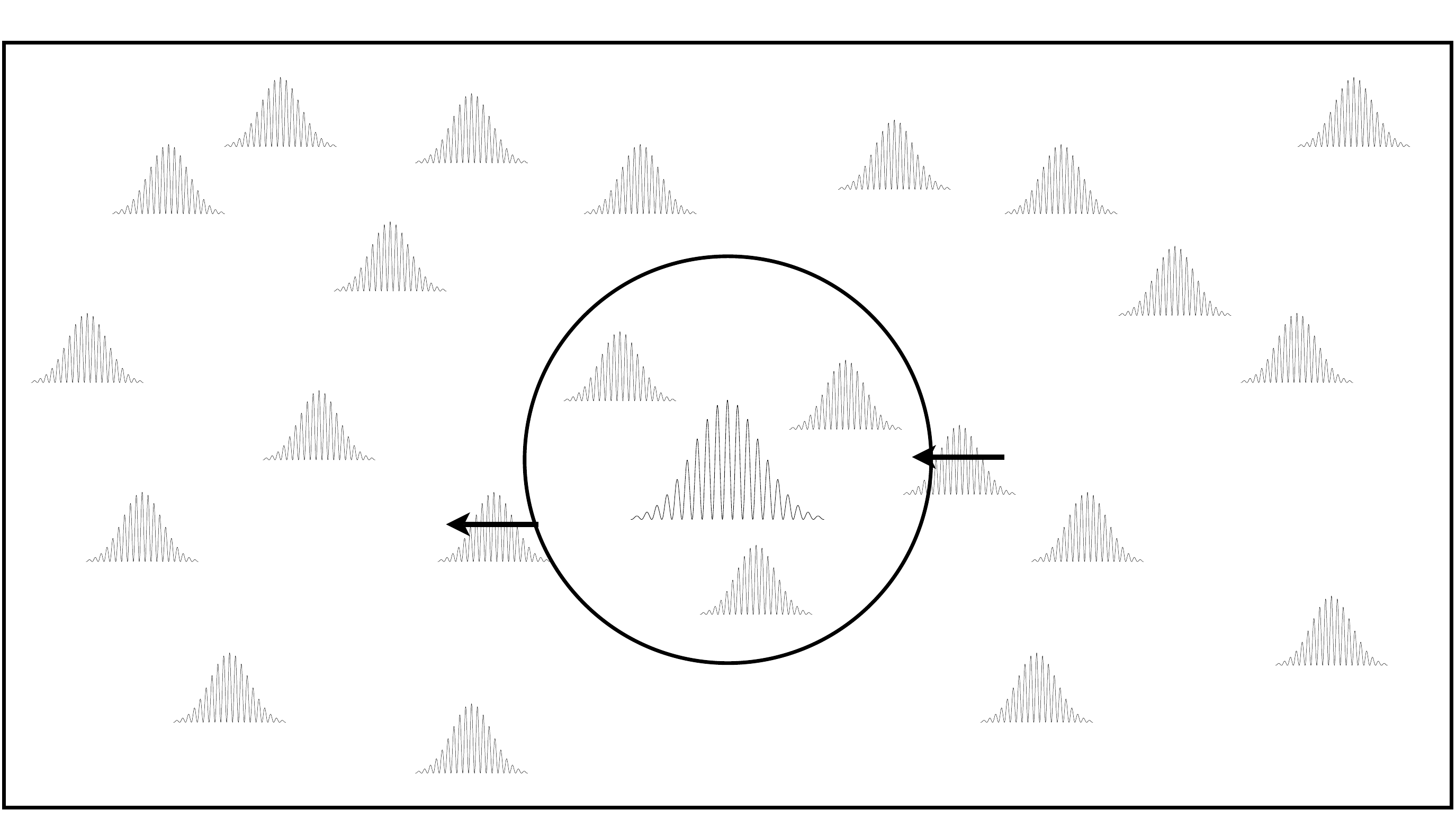}
\caption{Schematic representation of the system particle (large wave packet) with a Gaussian wave function
interacting with its environment (small wave packets).
The dynamics that is relevant for the system can be restricted to the vicinity around the particle
sketched by the circle.
The environment particle that leaves the vicinity to the left can be traced over since its impact on the
dynamics of the system particle is negligible.
On the other hand, there is another environment particle approaching the system particle from the right,
and needs to be taken into account.}
\label{fig1}
\end{figure}

We make use of the expansion into {\em localized} states which permits effectively reducing the
size of the environment to be considered. 
The underlying idea is to expand the state of the entire system (that is system and environment)
in terms of Gaussian states,
such that the actual dynamics can be reduced to those environment components that are situated in close proximity
to the system particle.
This allows  elimination of all other environmental degrees of freedom from the problem.
On the other hand, there will be environment particles that approach and enter this vicinity,
and their degrees of freedom will need to be incorporated in the dynamics
as sketched in Fig.~\ref{fig1}.

Any Gaussian state is completely characterized in terms of the expectation values of all coordinates and momenta
${\bf x}=[r_1,p_1,r_2,p_2,\hdots,r_n,p_n]=\tr\ {\bf \hat x}\varrho$ ,
and the corresponding covariances
\be
\Sigma_{ij}=\tr\ \frac{{\bf \hat x_i}{\bf\hat x_j}+{\bf\hat x_j}{\bf\hat x_i}}{2}\varrho\ -\ {\bf x}_i{\bf x}_j
\ee
However, an explicit parametrization of a Gaussian density matrix $\varrho$ in terms of those parameters
is a rather lengthy expression.
The Weyl symbol of the density matrix, {\it i.e.} the Wigner function
\be
W(\vec r,\vec p)=\int d^nq \bra{\vec r-\frac{\vec q}{2}}\varrho\ket{{\vec x+\frac{\vec q}{2}}}e^{\frac{i}{\hbar}\vec p\vec q}
\ee
is more convenient to use, since it has a significantly simpler parametrization:
\be
W(x)=\frac{1}{\sqrt{\pi^n\det\Sigma}}e^{-\frac{1}{2}\left(x-{\bf x}\right)\Sigma^{-1}\left(x-{\bf x}\right)}\ .
\label{wigner}
\ee
The evolution of a Gaussian quantum state due to a quadratic Hamiltonian
gives rise to Newton's equations of motion
$\dot{\bf x}={\cal S}\nabla H$ with the symplectic matrix {\cal S}, and
\be
\label{eqomreal}
\frac{\partial\Sigma}{\partial t}=
2(\Sigma{\cal H}{\cal S}-{\cal S}{\cal H}\Sigma)\ ,
\ee
where ${\cal H}$ contains the second derivatives of the Hamiltonian $H$ with respect to the coordinates
and momenta, {\it i.e.} ${\cal H}_{ij}=\frac{1}{2}\frac{\partial^2 H}{\partial{\bf x}_i\partial{\bf x}_j}$
\begin{figure}
\includegraphics[width=0.45\textwidth]{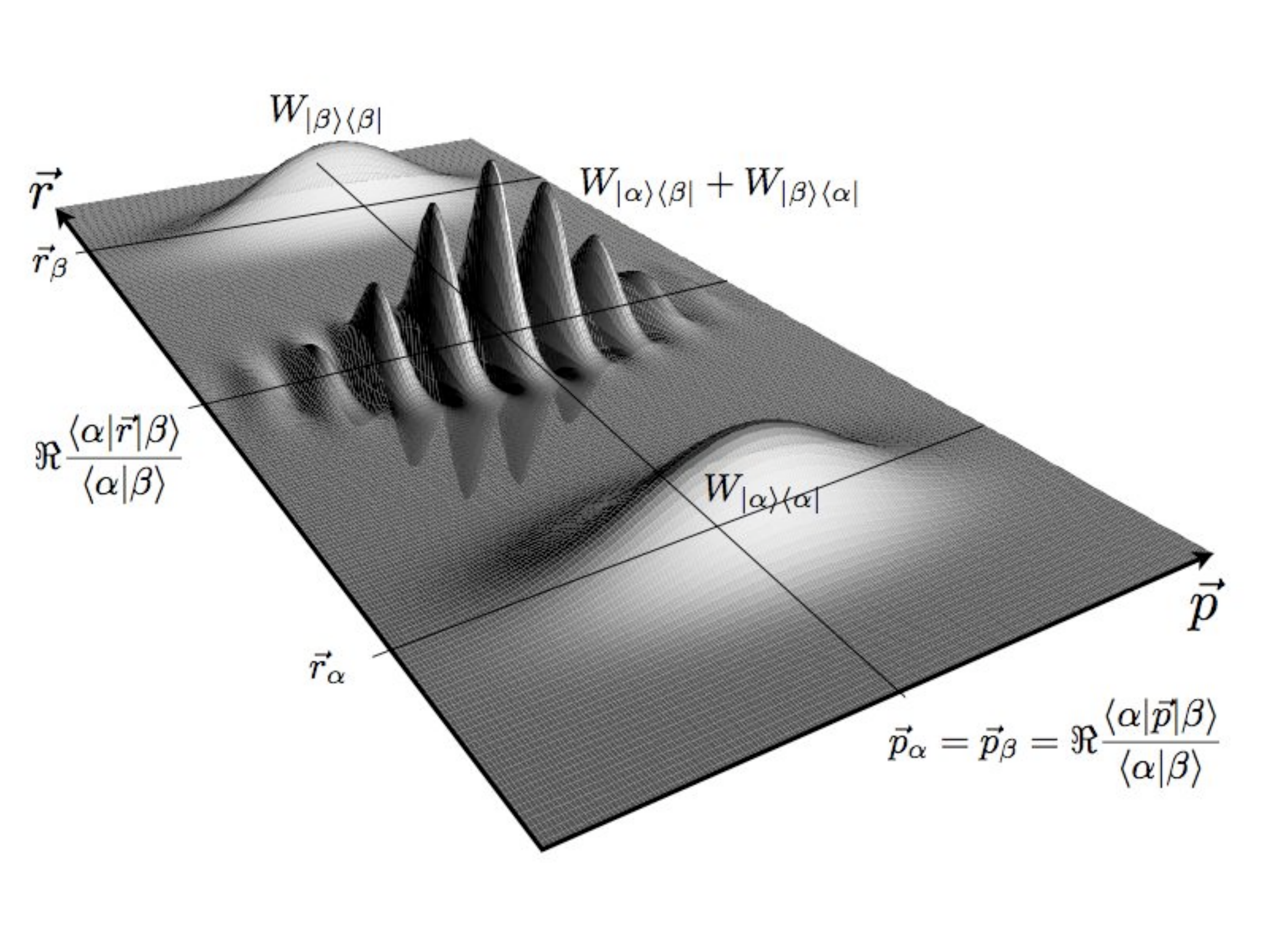}
\caption{The Wigner function for the superposition of two Gaussian states $\ket{\alpha}+\ket{\beta}$
is comprised of four parts.
The two `diagonal' parts corresponding to $\ket{\alpha}$, and $\ket{\beta}$ respectively,
are proper Wigner functions on their own, and their evolution is described by Eq.~\eqref{eqomreal}.
The Weyl symbol for the two `off-diagonal' parts corresponding to $\ket{\alpha}\bra{\beta}$, and $\ket{\beta}\bra{\alpha}$ are in general complex,
but they can be propagated similary to a regular Wigner function.
However, their dynamics is {\em not} governed by the Hamiltonian around the phase space position,
at which it is located, but around the positions of both $\ket{\alpha}$, and $\ket{\beta}$.}
\label{fig2}
\end{figure}

Eq.~(\ref{eqomreal}) describes the {\em unitary} evolution of the system and its surrounding environment.
Any dissipative nature of the the dynamics is due to tracing over environmental degrees of freedom.
Obviously, the expectation value of any observable $A$ on particles that are not being traced over
are unaffected by the partial tracing
\be
\tr A\rho=\tr A(\tr_p\rho)\ .
\ee
Therefore, tracing over some degrees of freedom is equivalent to simply dropping all components of
${\bf x}$ and $\Sigma$ corresponding to the degrees of freedom  being traced over. 
In turn, adding a particle corresponds to extending ${\bf x}$ and $\Sigma$ by the respective quantities,
with vanishing correlations between the added particle, and the residual system.

So far, we have been assuming the initial state be a spatially narrow Gaussian wave packet.
This condition, however, is often not satisfied,
and for wider wave packets the harmonic approximation breaks down.
Nevertheless, {\em any} quantum state $\varrho$ (pure or mixed) can be decomposed into an incoherent mixture
of coherent states
$\varrho=\int d\mu(\alpha)\ P_\alpha(\varrho)\ \ket{\alpha}\bra{\alpha}$,
with the $P$-function $P_\alpha(\varrho)$.
Since, this holds also for pure states, $P_\alpha$ cannot always be a probability distribution,
{\it i.e.} it can adopt negative values.
The central merit of this representation is that different initial states $\ket{\alpha}$ can be propagated
individually,
and the overall final state is then the incoherent sum over the individually propagated states.
In the presently discussed framework of open system, this implies that partial traces can be taken at any instance.
The big disadvantage of such a representation is the often wild behavior of $P_\alpha$.
In particular for non-classical states $P_\alpha$ is rapidly oscillating, and close to singular,
what severely limits its usefulness for practical purposes.

Often it is significantly easier to expand the state $\ket{\Psi}$ into a coherent sum of Gaussian states
$\ket{\Psi}\simeq\sum_\alpha\Psi_{\alpha}\ket{\alpha}$.
Due to the linearity of the Schr\"odinger equation,
one can of course propagate each initial term $\ket{\alpha_i}$ individually,
and, then reconstruct the final state
\be
{\cal U}_t\ket{\Psi}=\sum_\alpha\Psi_\alpha{\cal U}_{t}\ket{\alpha}\simeq\sum_\alpha\Psi_\alpha{\cal U}_t^{\alpha}\ket{\alpha}\ ,
\ee
where ${\cal U}_t^{\alpha}=T\exp(-i/\hbar \int dt H_\alpha)$ is the approximate propagator,
and $H_\alpha$ is the locally quadratic approximation of $H$ expanded around ${\bf x}_{\alpha}$.

In order to take partial traces, however, one needs to consider the corresponding density matrix:
\be
\varrho_t=\sum_{\alpha\beta}\Psi_\alpha\Psi_\beta^\ast\ {\cal U}_t\ket{\alpha}\bra{\beta}{\cal U}_t^\dagger\simeq
\sum_{\alpha\beta}\Psi_\alpha\Psi_\beta^\ast\ {\cal U}_t^{\alpha}\ket{\alpha}\bra{\beta}{\cal U}_t^{\beta^\dagger}\ .
\ee
Here, ${\cal U}_t^{\alpha,\beta}$ is the unitary transformation generated by the Hamiltonian $H$ expanded around the
classical positions of the states $\ket{\alpha}$ and $\ket{\beta}$ respectively.
Thus, in the following, we will consider the evolution of the individual operators
$\rho_{\alpha\beta}$ resulting from $\ket{\alpha}\bra{\beta}$.
Since those operators are not necessarily normalized,
all expectation values will be defined including normalization
$\langle A\rangle_\rho=\tr A\rho/\tr\rho$.
Doing so, one obtains the equations of motion
\be
\label{eqomcomplex}
\frac{\partial\Sigma}{\partial t}=
2(\Sigma{\cal H}_+{\cal S}-{\cal S}{\cal H}_+\Sigma)-
\frac{i\hbar}{2}{\cal S}{\cal H}_-{\cal S}-\frac{2i}{\hbar}\Sigma{\cal H}_-\Sigma\ ,
\ee
with
${\cal H}_+=1/2({\cal H}_{\alpha}+{\cal H}_{\beta})$, and
${\cal H}_-={\cal H}_{\alpha}-{\cal H}_{\beta}$;
and with ${\cal H}_{\alpha}$, ${\cal H}_{\beta}$ being the second order expansion coefficients of $H$
taken along the phase space positions $\bra{\alpha}\hat{\bf x}\ket{\alpha}$, $\bra{\beta}\hat{\bf x}\ket{\beta}$
as depicted in Fig.~\ref{fig2}.
The equations of motion for the positions and momenta ${\bf x}$ are a bit more lengthy,
and they are most conveniently characterized by the relation
\be
\label{eqomcomplex}
\vec {\bf x}=\frac{1}{2}(\vec {\bf x}_{\alpha}+\vec {\bf x}_{\beta})-\frac{i}{\hbar}\Sigma{\cal S}(\vec {\bf x}_{\alpha}-\vec {\bf x}_{\beta})\ ,
\ee
and
${\bf x}_\alpha$ and ${\bf x}_\beta$ evolving according to Newton's equation of motion.

The Weyl symbol corresponding to those complex phase space coordinates and uncertainties is given by Eq.~(\ref{wigner})
up to the additional factor $\exp(\eta)$ with
\be
\eta=\frac{1}{2\hbar^2}(\vec {\bf x}_{\alpha}-\vec {\bf x}_{\beta}){\cal S}\Sigma{\cal S}(\vec {\bf x}_{\alpha}-\vec {\bf x}_{\beta})-
\frac{i}{2\hbar}(\vec p_{\alpha}+\vec p_{\beta})(\vec r_{\alpha}-\vec r_{\beta})\ ,
\label{eqeta}
\ee
and a phase $\varphi$ whose evolution is given by
$\dot\varphi={\cal L}_{\alpha}-{\cal L}_{\beta}-\frac{1}{2}\tr\Sigma{\cal H}_-$,
where ${\cal L}_i$, ($i=1,2$) is the Lagrange function with variables $\vec r_i$, and $\vec p_i$.

\begin{figure}
\includegraphics[width=0.3\textwidth,angle=270]{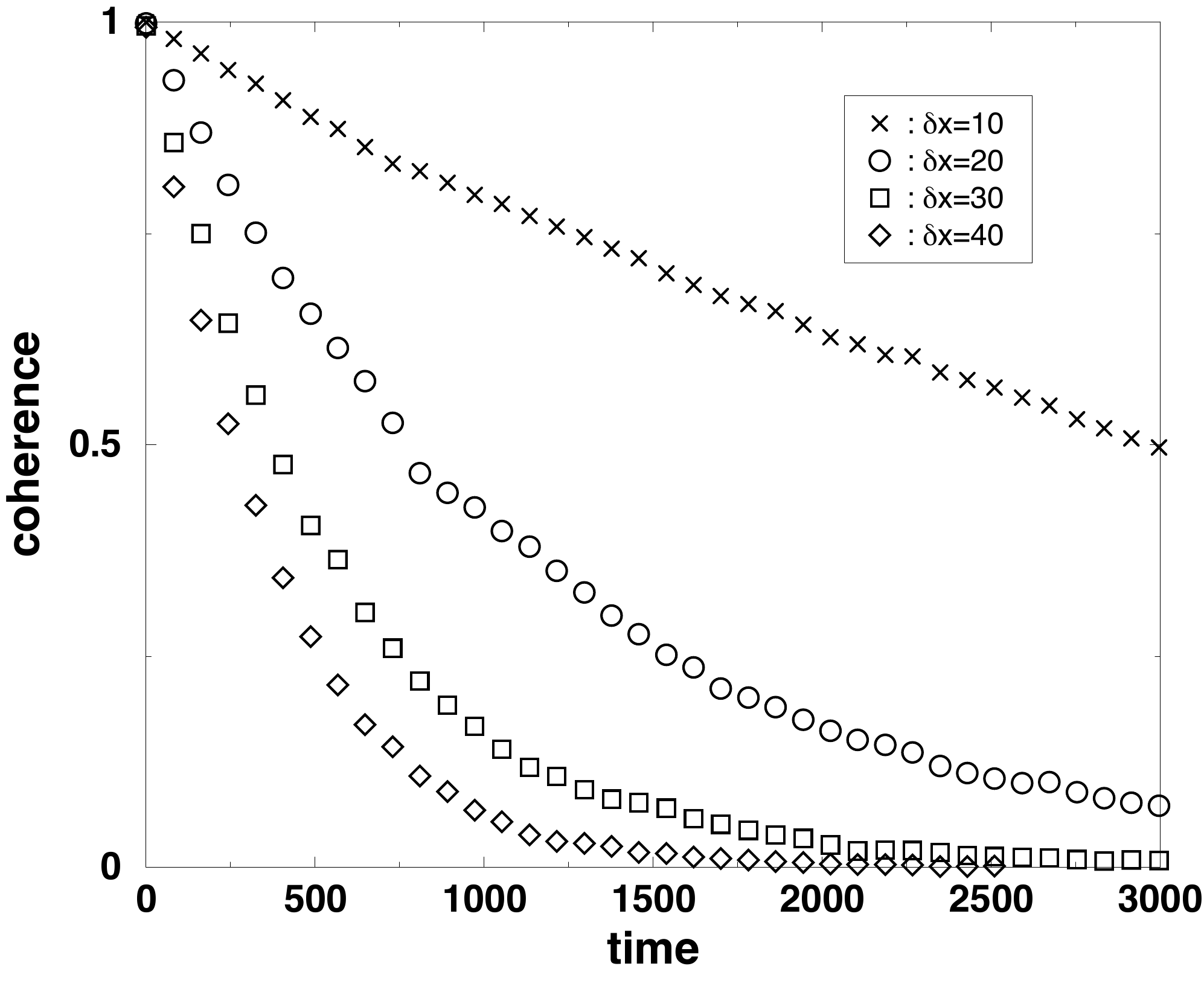}
\caption{The coherence of the superposition of $\ket{\alpha}$ and $\ket{\beta}$ gets reduced in time due to interactions
with an environment gas.
Here, this coherence is characterized in terms of the Hilbert-Schmidt norm of the operator that the initial operator $\ket{\alpha}\bra{\beta}$
evolves into,
and it is plotted as a function of time for $4$ different initial states, where the relative phase space separation $\delta x$ between $\ket{\alpha}$
and $\ket{\beta}$ is chosen to be $10$, $20$, $30$, and $40$ oscillator length.
The coherence decays exponentially, and the decays is faster for larger initial phase space separations.}
\label{fig3}
\end{figure}
So far we were concerned mainly with the unitary dynamics of the system.
However, some care has to be taken while tracing over an environmental particle.
As mentioned before, the dynamical variables ${\bf x}$, and $\Sigma$
remain unchanged under partial tracing.
However, ${\bf x}_\alpha$, and ${\bf x}_\beta$ are propagated rather than ${\bf x}$,
and for a general state any entry of ${\bf x}$ depends on all the entries of ${\bf x}_\alpha$, and ${\bf x}_\beta$
as shown in Eq.~\eqref{eqomcomplex}.
Therefore, one may not simply drop the entries of ${\bf x}_\alpha$, and ${\bf x}_\beta$ that belong to a particle
that is traced over,
but one needs to consider their contributions to ${\bf x}$ first.
More explicitly, after a partial trace has been performed, ${\bf x}$ is not given by Eq.~\eqref{eqomcomplex} anymore,
but there is the additional term
$i / \hbar\Sigma{\cal S}P({\bf x}_\alpha-{\bf x}_\beta)$
where $P$ is the projector onto the the space spanned by the phase space coordinates of all the particles that have been traced over.
Similarly, also $\eta$ in Eq.~\eqref{eqeta} depends on variables associated with particles that are being traced over.
Therefore, also those contributions have to be recorded explicitly during the removal of particles from the system.

In order to illustrate the present method we will apply it to the investigation of decoherence rates
of superpositions of harmonic oscillator coherent states $\ket{\alpha}+\ket{\beta}$.
The decoherence rate is predicted \cite{PhysRevA.31.2403,PhysRevA.31.1059} and experimentally verified \cite{PhysRevLett.77.4887,Myatt:2000lr}
to grow quadratically with increasing phase-space separation $\delta {\bf x}$ between $\ket{\alpha}$ and $\ket{\beta}$.
However, this quadratic dependence holds only for small phase space separations,
and once the separation is much larger then the range of the interaction potential a saturation is expected \cite{hornberger012105,klh}.
We consider the full 3-dimensional problem with a thermal environment of mutually noniteracting particles
that, however, interact with the oscillator via a short range Gaussian interaction.
The environment is dilute so that only two-body interactions are taken into account.
A thermal environment is realized via an average over $2000$ realizations of an environment consisting of $1500$ particles.
Due to the possibility of adding and removing particles this requires the integration of $182$ coupled differential equations,
whereas the simulation of the entire system with all particles present during the entire integration would yield $8 \times 10^7$ differential equations.

\begin{figure}[t]
\includegraphics[width=0.3\textwidth,angle=270]{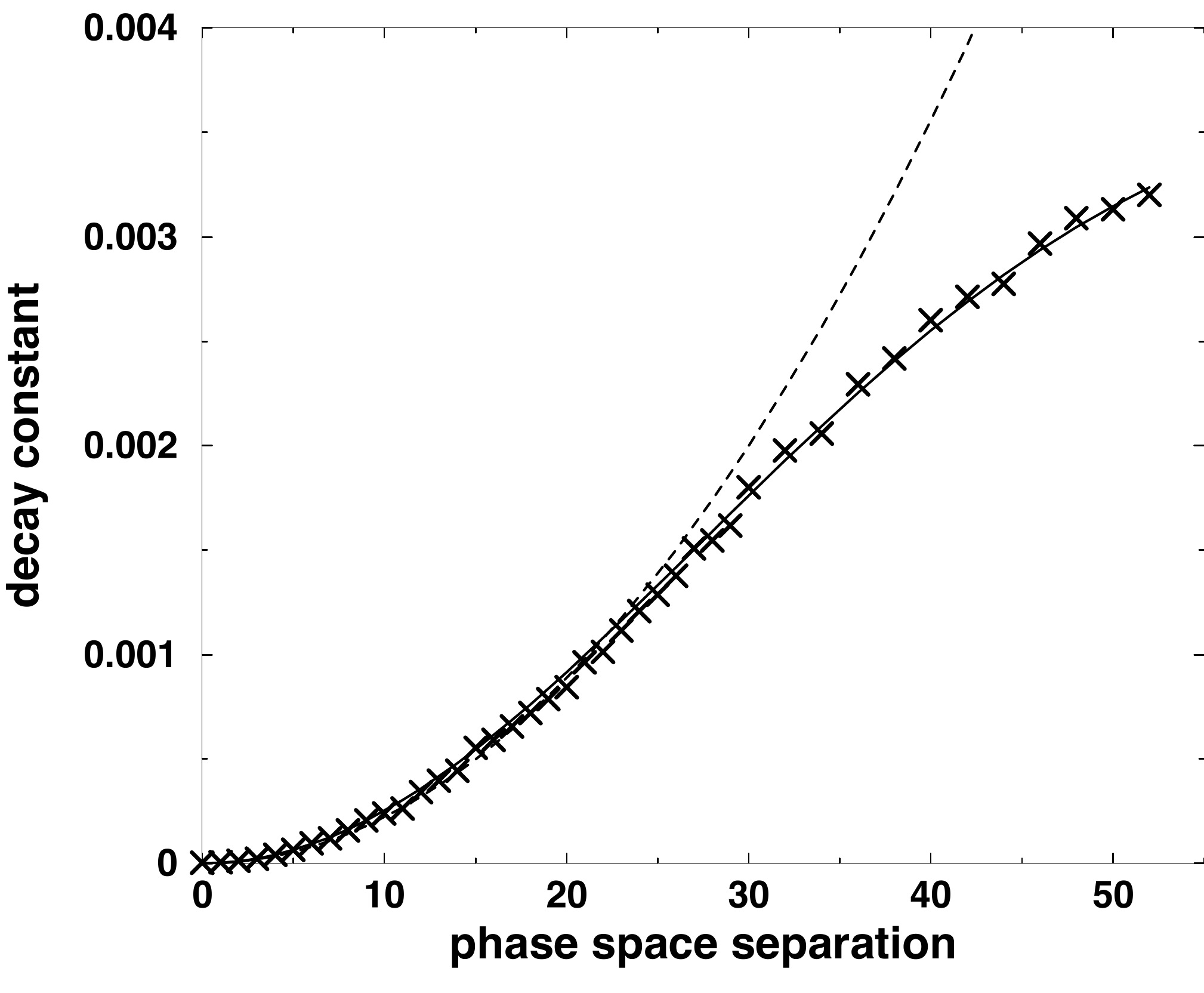}
\caption{Decay constants for the coherence as in Fig.~\ref{fig3} as function of the initial relative phase space separation $\delta {\bf x}$ between $\ket{\alpha}$ and $\ket{\beta}$.
For small $\delta x$ he decay constants grow quadratically with the phase space separation.
Starting with $\delta z$ exceeding the interaction range ($25$ oscillator length), a deviation from the quadratic growth (dashed line) sets in,
and the behavior is described better by a Gaussian (solid line).}
\label{fig4}
\end{figure}

Fig.~\ref{fig3} shows the decay of the Hilbert Schmidt norm $\tr\varrho_{\alpha\beta}\varrho_{\alpha\beta}^\dagger$,
{\it i.e.} the magnitude of coherence as function of time for different initial phase space separations $10$, $20$, $30$ and $40$ oscillator lengths.
The individual curves show an exponential decay $\exp(-\gamma t)$ (with time given in multiples of the oscillator period)
with some remaining noise due to the average over a finite sample of environment realizations,
and allow the extraction of a decay constant $\gamma$.
Fig.\ref{fig4} displays this decay constant as function of the initial phase space separation.
Even for $\delta {\bf x}$ vanishing, there is some decoherence with a decay constant of about $10^{-5}$ since a coherent state is not an eigenstate to the interaction potential,
but the decay time for this case is significantly longer than that for a coherent superposition of two coherent states.
For small values of $\delta {\bf x}$, we recover the predicted quadratic increase of the decay constant,
and once the phase space separation exceeds the range of the interaction potential ($25$ oscillator lengths in this case),
the increase with $\delta {\bf x}$ gets slower and, as expected, saturation sets in.

The present techniques allow  treatment of  general situations of quantum systems embedded in an environment.
The range of applicability is basically limited only by the system-potentials,  which should vary slowly over the typical width
of the wave packets that are used.
Since wave packets spread spatially over  time in many systems, there comes a  time after which the
present approximations become questionable.
Nevertheless, since decoherence typically takes place on time-scales that are significantly shorter than other damping phenomena,
it is rather the short-term than long-term behavior that is of importance for open quantum systems.
It is also important to note in this context that an environment interaction can result in a narrowing of the system wave packet, so that with an environment  there is an effect  counteracting  the  spreading.

We would like to thank Thierry Paul and Thomas Pohl for stimulating discussions.
Financial support by the Alexander von Humboldt fundation is gratefully acknowledged.

\bibliography{../../../Referenzen}

\end{document}